\documentclass[Transaction, 9pt]{IEEEtran}
\usepackage{times}
\usepackage{graphicx}
\usepackage[top=2.3cm,bottom=2.3cm,inner=1.5cm,outer=1.5cm,bindingoffset=0cm]{geometry}
\begin{document}
%\title{Polarity Control Doping-Less Tunnel Field Effect Transistor}
\title{A Dynamically Configurable Silicon Nanowire Field Effect Transistor based on Electrically Doped Source/Drain}
\author{ Chitrakant~Sahu, Avinash~Lahgere and ~Jawar~Singh\\
% <-this % stops a space
{PDPM Indian Institute of Information Technology Design and Manufacturing, Jabalpur, MP, India} \\
E-mail: {(chitrakant, avinash.lahgere, jawar@iiitdmj.ac.in).}}
\maketitle

%\begin{abstract}

%\end{abstract}

%\begin{IEEEkeywords}
%CMOS, MOSFETs, TFETs and configurable FETs.
%\end{IEEEkeywords}

%{\emph{Introduction:}}
In this article, we present a configurable field-effect transistor (FET), where not only polarity (n- and p-type), but the conduction mechanism of a FET can also be configured dynamically. As a result, we can have both types of devices, high-performance MOSFET and low-power TFET, for computational and power efficient system on chip (SoC) products. The calibrated 3D-TCAD simulation results validate characteristics and functionalities of the configurable FET, and showed good consistency with the static conventional MOSFET (or TFET). The device consist of uniformly and lowly doped ultrathin silicon layer from source to drain, where, carrier concentrations (electron or hole) are precisely controlled by appropriate application of external bias, instead of conventional doping in source/drain (S/D) region. Hence, the utilization of lowly doped regions offer less susceptibility to random dopant fluctuations (RDF) and absence of abrupt doping profile at the S/D leads to reduced fabrication complexity and low thermal budget~\cite{jag,ss}. Recently, configurable logic gates using polarity controlled silicon nanowire (SiNW) FET have been demonstrated with configurable n- and p-MOSFETs ~\cite{dem,demm}, using extra polarity gates and showed good potential for higher packaging density and compatibility with CMOS process.

\begin{figure}
\center
%\begin{floatrow}%
{  \includegraphics[width=0.35\textwidth]{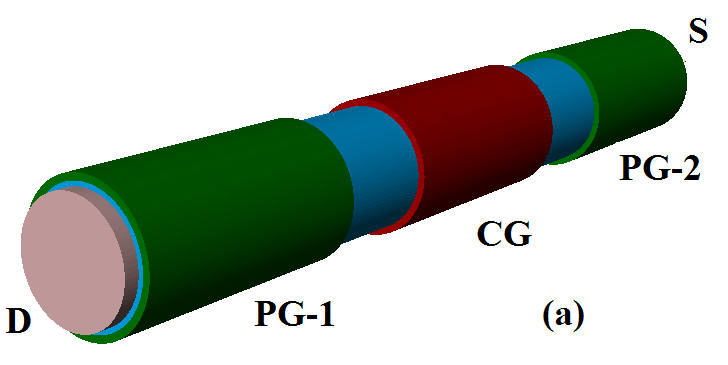}%
   \\ \includegraphics[width=0.43\textwidth]{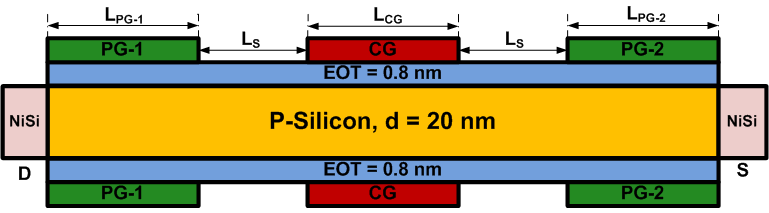}%
   }{\caption{(a) Schematic structure and (b) cross section of n-type circular SiNW configurable FET.}\label{device}}%
%\end{floatrow}%
\end{figure}

Here, we have showed that the electrically doped S/D regions of a device allows it to operate as a MOSFET as well as TFET, not just a n- and p-MOSFETs, referred as configurable FET. Fig.~\ref{device} (a) and (b) show 3D and cross sectional view of configurable FET, respectively. It consist of two sets of gate electrodes: (a) the control gate (CG) acts identically as conventional gate for switching ON and OFF, and vice versa, and (b) the polarity gates (PGs) embedded on the side regions of the channel close to source and drain contacts, they control's the polarity and conduction mechanism of the device. The source and drain regions (either $P^{+}$ or $N^{+}$ ) are induced on ultrathin lowly doped silicon body by applying external bias to PGs. Further, source and drain contacts are made up of nickel silicide (NiSi) with barrier height 0.45 eV. The working mechanism of configurable FET in (a) MOSFET mode and (b) TFET mode is illustrated by various contour plots, band diagram and transfer characteristics using different models in ATLAS TCAD simulator~\cite{sil}. Due to difference in conduction mechanism of both devices, we have incorporated different models. For MOSFET, hydrodynamic transport model is enabled~\cite{dem}, whereas, for TFET the non-local band-to-band tunneling model is enabled in order to account tunneling mechanism~\cite{jag,cali}.

The carrier concentration contours for configurable FET in both (a) MOSFET and (b) TFET mode under OFF- and ON-state conditions are shown in Figs.2-4. One can observe that by employing different polarity gate biasing the configurable FET switch into either MOSFET to TFET or vice versa, when no control gate voltage is applied. For instance, if both PG-1 and PG-2 are set to 1.2 V then resulting device will behave like an MOSFET due to $N^{+}-I-N^{+}$ architecture in OFF state as shown in Fig. 2(a). However, it can be configured as TFET when PG-1 is set to 1.2 V and PG-2 is biased with -1.2 V that results in $N^{+}-I-P^{+}$ architecture can be seen in Fig. 3(a-b). The both device can be turned ON by control gate voltage due to inversion of carriers below control gate as shown by carrier concentration in Fig. 2(b) (MOSFET mode) and Fig. 4(a-b) (TFET mode). Further, OFF- and ON-state band diagram for configurable FET in MOSFET and TFET mode are shown in Fig.~\ref{band_mos}(a) and (b), respectively, it can seen that appropriate biasing of polarity and control gates distinctively exhibit the band bending and confirm the functionality of configurable MOSFET and TFET.

Fig.~\ref{idvg_both}(a) and (b) show the transfer characteristic of  configurable FET in MOSFET and TFET mode, respectively at different polarity bias conditions. It can be seen that higher PGs bias causes enhancement in the ON-state ($I_{ON}$) current for both devices due to accumulation of larger carrier concentration in source/drain regions. The configurable FET  yields a steep subthrehold swing (approx. 24 mV/dec) and significantly very low off-state current (order of $10^{-19}$ A) in n-TFET mode. However, it shows slightly low on-state current as compared to a device when it is configured as MOSFET. It happens because in TFET on-state ($I_{ON}$) current is governed by effective mass and band gap of material~\cite{jag}, whereas in MOSFET the on-state ($I_{ON}$) current is governed by majority carriers, which results in ON current of 35$\mu$A. The important factor for high performance device is ON current, whereas for low power device OFF current play a major role. So, we calculated ratio of $I_{ON}$ (MOSFET) to  $I_{OFF}$ (TFET) for a configurable FET, which is in the order of $10^{13}$. Hence, the configurable device achieves indispensable behavior dynamically with polarity gates, which is consistent with the static conventional MOSFET (TFET) and a suitable candidate for high-performance and low-power applications. However, proposed device also have a drawback of extra polarity gates, hence, extra metal deposition and isolation is required that may cause more parasitic effects.

\begin{figure}[ht]
\center
\includegraphics[height=35mm,width=40mm]{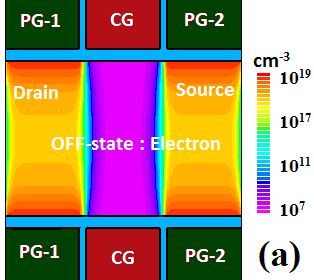}
 \includegraphics[height=35mm,width=40mm]{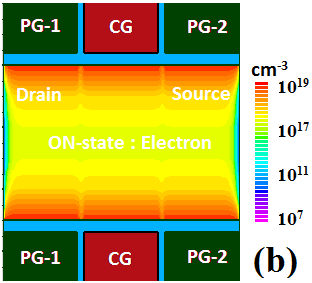}
\caption{Electron concentrations contour in (a) OFF-state ($V_{CG}$=0V, $V_{PG-1}$=$V_{PG-2}$=1.2V, and $V_{DS}$=0V) and (b) ON-state ($V_{CG}$=1.2V, $V_{PG-1}$=$V_{PG-2}$=1.2V, and $V_{DS}$=1.2V) conditions for configurable FET in MOSFET mode.}
\label{nmos}
\end{figure}

\begin{figure}[ht]
\center
\includegraphics[height=35mm,width=40mm]{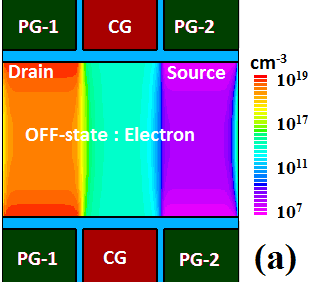}
  \includegraphics[height=35mm,width=40mm]{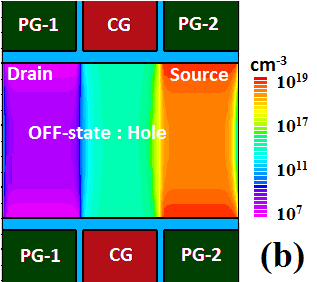}
\caption{Contour plots of (a) electron and (b) hole carrier concentrations in OFF-state ($V_{CG}$=0V, $V_{PG-1}$=1.2V, $V_{PG-2}$=-1.2V, and $V_{DS}$=0V) conditions for configurable FET in TFET mode.}
\label{ntfet_off}
\end{figure}

\begin{figure}[ht]
\center
\includegraphics[height=35mm,width=40mm]{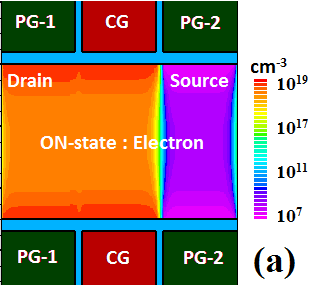}
 \includegraphics[height=35mm,width=40mm]{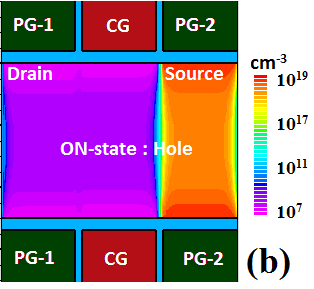}
\caption{Contour plots of (a) electron and (b) hole carrier concentrations in ON-state ($V_{CG}$=1.2V, $V_{PG-1}$=1.2V, $V_{PG-2}$=-1.2V and $V_{DS}$=0.5V) conditions for configurable FET in TFET mode.}\label{ntfet}
\end{figure}

\begin{figure}[ht]
\center
\includegraphics[width=70mm,keepaspectratio]{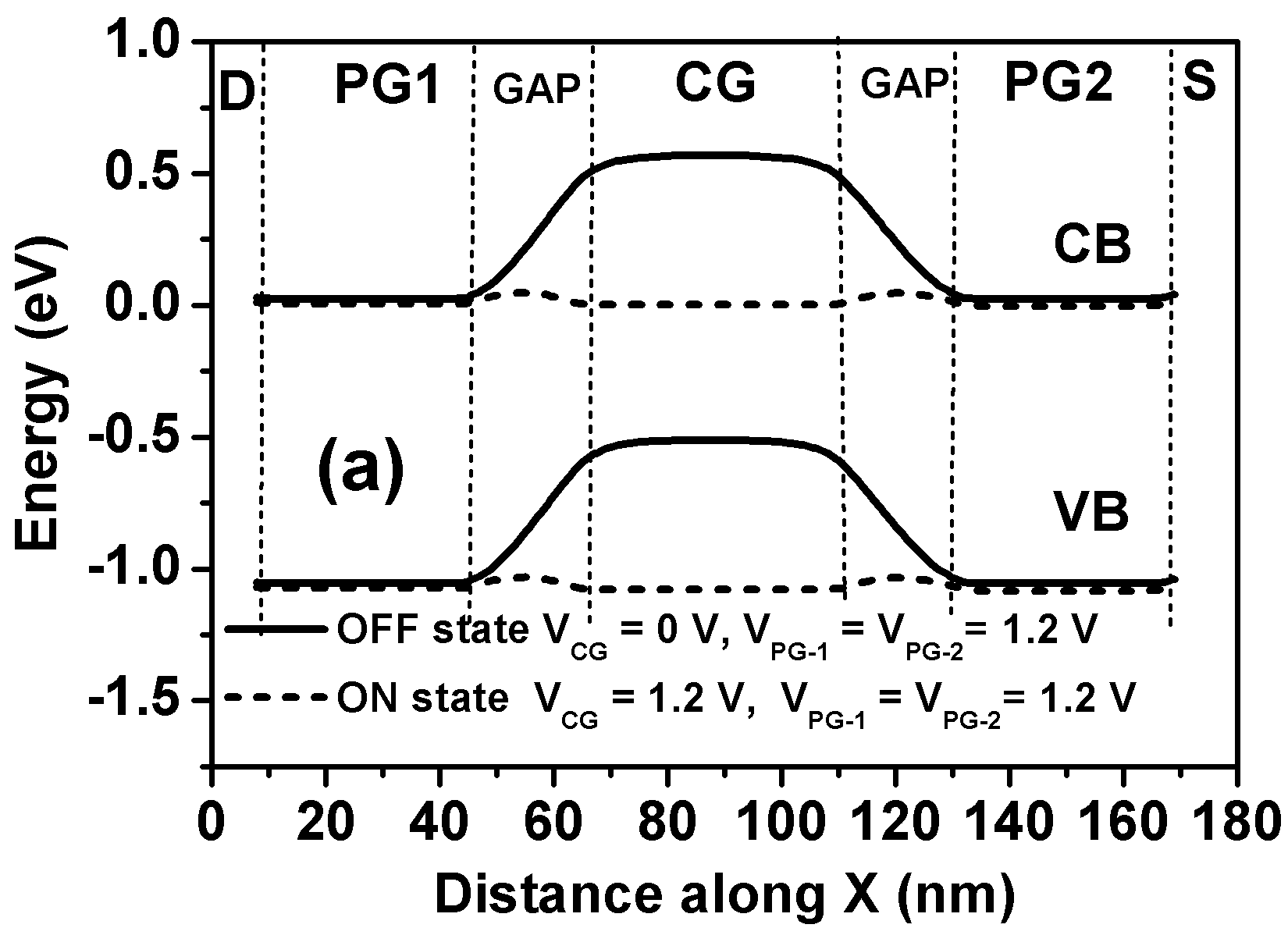}
\includegraphics[width=70mm,keepaspectratio]{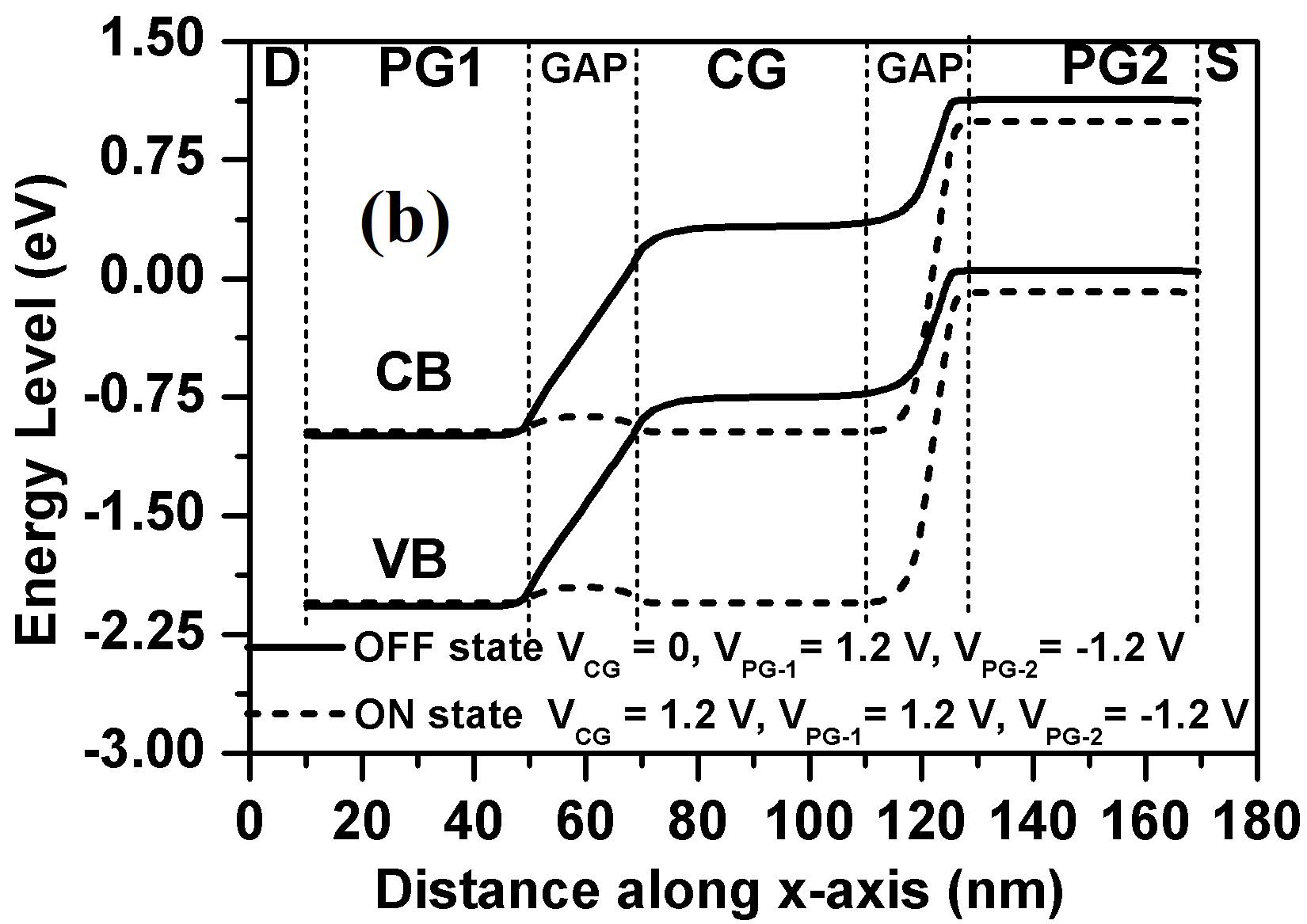}
\caption{OFF-state and ON-state band diagram of configurable FET in (a) MOSFET mode and (b) TFET mode.} \label{band_mos}
\end{figure}

\begin{figure}[ht]
\center
\includegraphics[width=70mm,keepaspectratio]{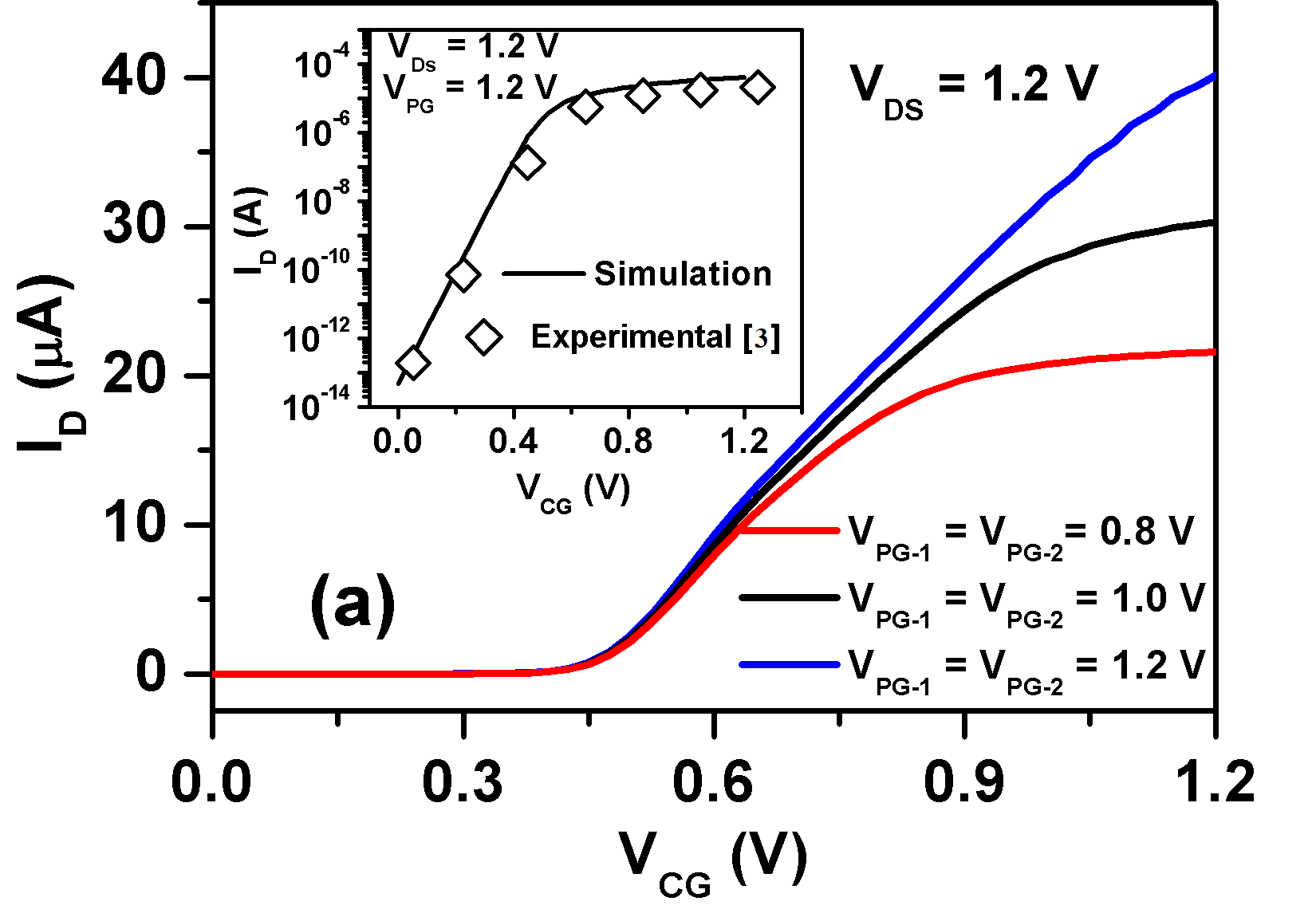}
\includegraphics[width=70mm,keepaspectratio]{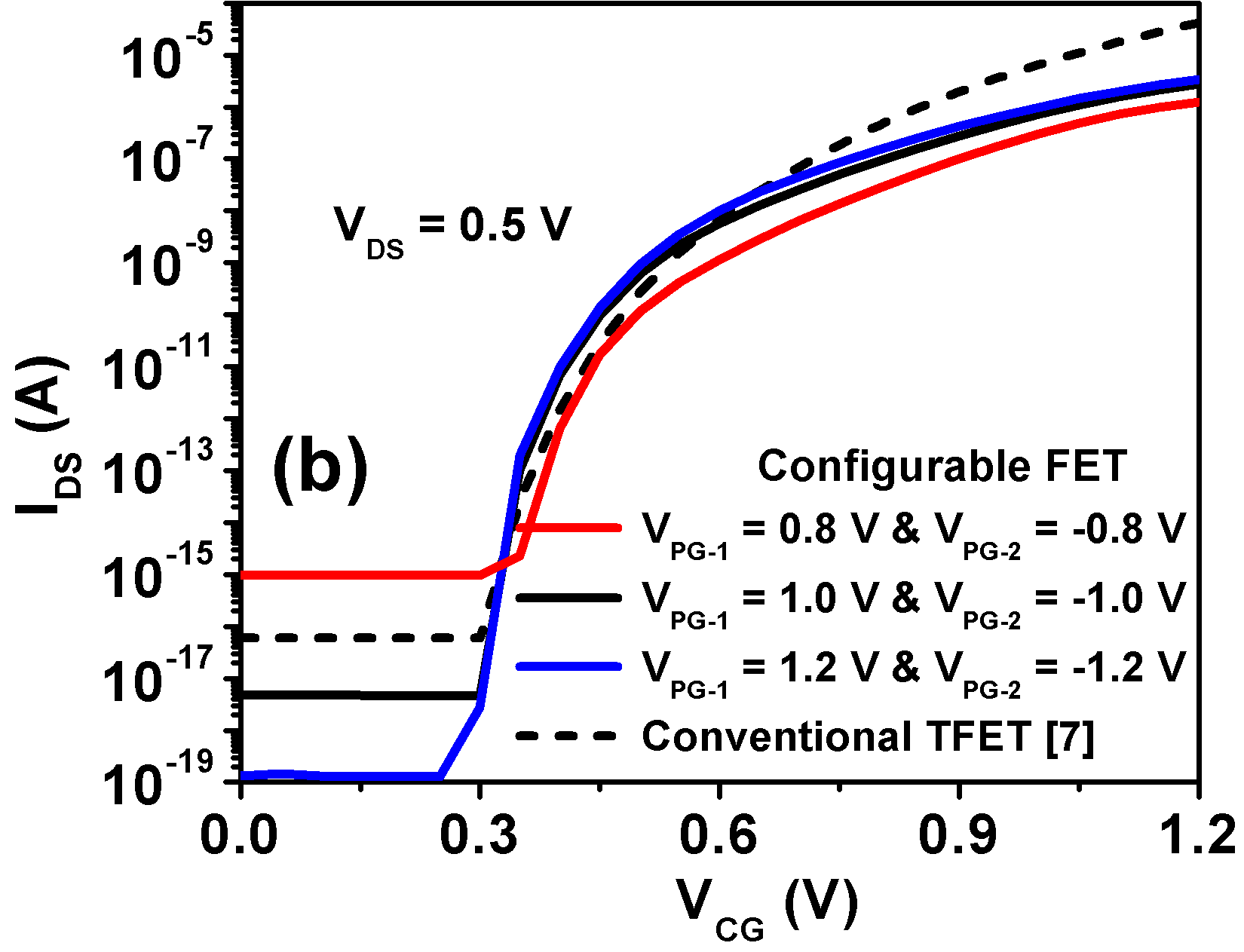}
\caption{Transfer characteristics of configurable FET in (a) MOSFET mode and (b) TFET mode along with calibration of simulation results with experimental data [3].} \label{idvg_both}
\end{figure}
%Fig.~\ref{loss}(a) and (b) show the static power loss ($P_{StaticLoss}=V_{DS}I_{Leakage}$) for n-, p-MOSFET (and TFET) at $V_{CG}$ = 0 V. Note that the static power loss is increasing with $V_{DS}$, but MOSFETs exhibit more dependence and higher 5-6 order of magnitude than TFETs. It is because of TFETs low probability of tunneling carriers, whereas, in MOSFETs lateral electric field increases more number of carriers to inject thermally. Fig.~\ref{loss}(c) and (d) show the intrinsic device delay ($\tau=\frac{C_{GG}V_{DD}}{I_{DS}}$) for MOSFETs and TFETs. One can observe that the $\tau$ decreases with increase in $V_{DS}$, it is due to enhanced $I_{ON}$ and reduced gate-to-gate capacitance ($C_{GG}$). However, $\tau$ for MOSFETs is lower 5-6 order of magnitude than TFETs, because of low $I_{ON}$ in TFETs as compared to their counterpart MOSFETs. Hence, same device can be configured to achieve high-performance (MOSFET) and power saving (TFET) modes.

%From power delay product ($PDP=\tau \times P_{Static Loss}$), which is also an essential performance metric for device selection. To this end, we observed the PDP of dynamically configurable MOSFETs and TFETs with $V_{DS}$, as shown in Fig.~\ref{pdp}(a) and (b). One can observe that the PDP for TFETs and MOSFETs is decreasing with $V_{DS}$ which is very obvious, as intrinsic device delay reduces significantly. However, PDP for TFETs is significantly lower than the MOSFETs, which is ideal for complementary circuit implementation.

%\section{Conclusion}

\end{document}